\documentclass{article}
\usepackage[margin=1in]{geometry}
\usepackage{amsmath,amssymb,graphicx,bm}
\usepackage{natbib}
\usepackage[colorlinks=true, allcolors=blue]{hyperref}
\usepackage{authblk}
\usepackage{float}
\usepackage[ruled]{algorithm2e}
\usepackage{xcolor}

\newcommand{\btheta}{\boldsymbol{\theta}}
\newcommand{\bx}{\mathbf{x}}
\newcommand{\dd}{\mathrm{d}}

\renewcommand{\mid}{\;\ifnum\currentgrouptype=16 \middle\fi|\;}

\title{Estimating Marginal Likelihoods in Likelihood-Free Inference via Neural Density Estimation}
\author[1]{Paul Bastide}
\author[2]{Arnaud Estoup}
\author[3]{Jean-Michel Marin}
\author[4, 5]{Julien Stoehr}
\affil[1]{Universit\'e Paris Cit\'e, CNRS, MAP5, F-75006 Paris, France}
\affil[2]{CBGP, INRAE, CIRAD, IRD, Montpellier SupAgro, Universit\'e Montpellier, Montpellier, France}
\affil[3]{IMAG, Universit\'e de Montpellier, CNRS, 34090 Montpellier, France}
\affil[4]{Universit\'e Paris-Dauphine, Universit\'e PSL, CNRS, CEREMADE, 75016 Paris, France}
\affil[5]{Université Paris-Saclay, AgroParisTech, INRAE, UMR MIA Paris-Saclay, 91120, Palaiseau, France}

\date{July, 2025}

\begin{document}

\maketitle

\begin{abstract}
The marginal likelihood, or evidence, plays a central role in Bayesian model selection, yet remains notoriously challenging to compute in likelihood-free settings. While Simulation-Based Inference (SBI) techniques such as Sequential Neural Likelihood Estimation (SNLE) offer powerful tools to approximate posteriors using neural density estimators, they typically do not provide estimates of the evidence. In this technical report presented at BayesComp 2025, we present a simple and general methodology to estimate the marginal likelihood using the output of SNLE. 
\end{abstract}

\section{Introduction}

\subsection{Context}

Bayesian inference provides a principled framework for learning from data by combining prior information with observed evidence. A central quantity in this framework is the \emph{marginal likelihood} or \emph{evidence}, defined as the model likelihood integrated over the prior distribution, namely
\begin{equation*}
    p(\bx) = \int f(\bx|\btheta) p(\btheta) \, \dd\btheta.
\end{equation*}
The latter serves as the normalizing constant in Bayes' theorem and underpins model comparison. However, computing this integral is often intractable, particularly in complex models where the likelihood $f(\bx|\btheta)$ is unavailable in closed form or prohibitively expensive to evaluate. Simulation-Based Inference (SBI), also known as likelihood-free inference, addresses this challenge by circumventing the need for an explicit likelihood function. In the typical setting, we assume a generative model $f(\bx \mid \btheta)$ from which simulations can be drawn for any parameter value $\btheta \in \Theta$, even though the likelihood function itself is not pointwise evaluable. A proper prior $\pi(\btheta)$ is specified, and the goal becomes to infer the posterior $p(\btheta \mid \bx^*)$ from simulations alone. One of the earliest and most influential families of methods in this space is Approximate Bayesian Computation (ABC), which proceeds by simulating pseudo-datasets and comparing them to the observed dataset via summary statistics and a tolerance criterion. Since the foundational work of \citet{Tavare1997} and \citet{Pritchard1999}, ABC has evolved significantly, often incorporating machine learning techniques to improve efficiency and robustness \citep{Pudlo2016, Sheehan2016}. More recently, the SBI landscape has been transformed by the introduction of neural network-based techniques, which leverage modern density estimation and amortized inference tools. These methods are typically categorized by the quantity they aim to approximate:
\begin{itemize}
  \item \textbf{NPE} and \textbf{SNPE} (Sequential Neural Posterior Estimation) targets the posterior directly;
  \item \textbf{NLE} and \textbf{SNLE} (Sequential Neural Likelihood Estimation) learns a surrogate for the likelihood;
  \item \textbf{NRE} and \textbf{SNRE} (Sequential Neural Ratio Estimation) focuses on likelihood ratios.
\end{itemize}
Notable contributions in this space include \citet{Papamakarios2016, Papamakarios2019, Greenberg2019, Hermans2020, Cranmer2020}, and a comprehensive benchmarking study by \citet{Lueckmann2021} provides a detailed comparison of these approaches.

While these neural SBI methods have significantly improved the quality of posterior inference, they are typically not designed to provide estimates of the marginal likelihood (with the notable recent exception of \citealt{Spurio2023}, see below). As a result, Bayesian model comparison, a difficult task in the likelihood-free setting, is dominated by ABC methodologies \citep{Grelaud2009,Pudlo2016,Marin2018}. In this work, we focus on addressing this gap by proposing a family of methods that leverage the output of SNLE to estimate the marginal likelihood
\begin{equation}
C = \int f(\bx^* \mid \btheta) \pi(\btheta)\dd\btheta,
\label{eq:evidence}
\end{equation}
for some observed data $\bx^*$, despite the intractability of $f(\bx^* \mid \btheta)$. 


\subsection{Sequential Neural Likelihood Estimation}

Sequential Neural Likelihood Estimation (SNLE) is a powerful approach within SBI that approximates the likelihood function $f(\bx \mid \btheta)$ using a neural density estimator. This approximation is constructed iteratively over $L$ rounds of adaptive simulation, refining the learned likelihood by focusing simulations in regions of high posterior probability (see Algorithm \ref{algo:snle}).
In this work, we focus specifically on density estimators based on normalizing flows (NF) \citep{Papamakarios2021}.


After completing all rounds, samples from the final approximate posterior are obtained via MCMC targeting $\widehat{\pi}^{(L)}(\btheta \mid \bx^*)$. SNLE has shown to match or surpass other SBI methods in terms of inference quality, while often requiring significantly fewer simulations. That said, SNLE requires careful tuning of its components—architecture of the density estimator, learning rate schedules, and sampling strategy for MCMC. It also inherits the sensitivity to model misspecification common in simulation-based approaches.

SNLE is a prominent method within simulation-based inference (SBI) that focuses on iteratively refining a surrogate likelihood using rounds of targeted simulations. Unlike fully amortized approaches such as Neural Posterior Estimation (NPE) or Neural Likelihood Estimation (NLE), which train a single model to generalize across all possible observations, SNLE concentrates computational effort on the region of parameter space relevant to a specific observed dataset \( \bx^* \). This sequential adaptation enables SNLE to achieve higher accuracy with fewer simulations, making it especially effective in settings where simulation is expensive.

Amortized inference methods like those proposed by \citet{Papamakarios2016} and further developed by \citet{Cranmer2020} aim to build models that generalize across multiple inference tasks. While this amortization offers efficiency for repeated queries, it may underperform in single-instance inference due to limited expressivity or training resource constraints. In contrast, SNLE trades amortization for focus: each round adapts the simulator budget to the posterior’s high-probability region, leading to more precise inference at the cost of reuse across datasets.

While SNLE was originally introduced as a method for posterior inference in likelihood-free settings, its outputs contain richer information that can be exploited beyond posterior approximation. In particular, SNLE produces a surrogate likelihood function $q^{(L)}(\bx \mid \btheta)$ and uses it in combination with the prior $\pi(\btheta)$ to construct an approximate posterior distribution.

In this work, we show that these components can also be used to estimate the marginal likelihood $C$. Indeed, although the true likelihood cannot be evaluated, SNLE provides a trained surrogate $q^{(L)}(\bx^* \mid \btheta)$ that approximates it in regions of high posterior mass. Our goal is to develop practical strategies for estimating the marginal likelihood $C$ using only this surrogate likelihood and the posterior samples generated during SNLE’s sequential training procedure. These strategies enable Bayesian model comparison in simulation-based contexts, where the marginal likelihood is otherwise inaccessible.

\begin{algorithm}[!t]
\DontPrintSemicolon
\SetAlgoNlRelativeSize{-1}
\KwIn{observed dataset $\bx^{*}$, prior distribution $\pi(\cdot)$, simulator $f(\cdot \mid \btheta)$, number of iterations $L$, number of simulations per iteration $N$}
\KwOut{Trained posterior estimator $\widehat{\pi}^{(L)}(\cdot \mid \bx^{*})$}

\BlankLine
Set initial proposal  $\widehat{\pi}^{(0)}(\cdot \mid \bx^{*}) = \pi(\btheta)$ and training dataset $\mathcal{D} = \emptyset$\;

\For{$\ell = 1$ to $L$}{
	\For{$i = 1$ to $N$}{
	    	Sample $\btheta^{(\ell)}_i \sim \widehat{\pi}^{(\ell - 1)}(\cdot \mid \bx^{*})$\;
    		Sample $\bx^{(\ell)}_i \sim f\left(\cdot \mid \btheta^{(\ell)}_i\right)$\;
    		Add $\left(\btheta^{(\ell)}_i, \bx^{(\ell)}_i\right)$ to $\mathcal{D}$\;
	}
    
    	\BlankLine
    	Train $q^{(\ell)}(\bx \mid \btheta)$ on dataset $\mathcal{D}$ using maximum likelihood\;
	Set $\widehat{\pi}^{(\ell)}(\cdot \mid \bx^{*}) \propto q^{(\ell)}(\bx^{*} \mid \btheta)\pi(\btheta)$\;
}

\Return  $\widehat{\pi}^{(L)}(\cdot \mid \bx^{*})$\;
\caption{SNLE \citep{Papamakarios2019}}
\label{algo:snle}
\end{algorithm}

\section{The SIS-SNLE formulation}
We introduce here a Sequential Importance Sampling (SIS) \citep{Owen2013,Robert2004} technique, that can leverage the iterative nature of SNLE to progressively build an estimate of the marginal likelihood.

Let $q^{(\ell)}(\bx^* \mid \btheta)$ denote the neural likelihood approximation obtained after round $\ell$ of SNLE. We can then define a surrogate for the evidence associated with this intermediate approximation at round $\ell$ as
\[
C_{\ell} = \int q^{(\ell)}(\bx^* \mid \btheta) \pi(\btheta) \dd\btheta.
\]
The final estimate $C_L$ after $L$ rounds is expected to closely approximate the true marginal likelihood \eqref{eq:evidence}.

We now introduce a SIS estimator for $C_L$, based on evaluating the ratio $C_{\ell} / C_{\ell-1}$ across rounds. Assuming the prior distribution is absolutely continuous with respect to the SNLE surrogate posterior $\widehat{\pi}^{(\ell-1)}(\btheta \mid \bx^*)$, a simple change of measure leads to
\begin{align*}
C_{\ell} &= \int q^{(\ell)}(\bx^* \mid \btheta) \pi(\btheta) \dd\btheta \\
    &= \int q^{(\ell)}(\bx^* \mid \btheta) \frac{\pi(\btheta)}{\widehat{\pi}^{(\ell-1)}(\btheta \mid \bx^*)} \widehat{\pi}^{(\ell-1)}(\btheta \mid \bx^*) \dd\btheta,
\end{align*}
where the posterior from the previous round satisfies
\[
\widehat{\pi}^{(\ell-1)}(\btheta \mid \bx^*) = \frac{\pi(\btheta) q^{(\ell-1)}(\bx^* \mid \btheta)}{C_{\ell-1}}\,.
\]
It follows that
\[
R_\ell = \frac{C_\ell}{C_{\ell-1}} = \int \frac{q^{(\ell)}(\bx^* \mid \btheta)}{q^{(\ell-1)}(\bx^* \mid \btheta)} \widehat{\pi}^{(\ell-1)}(\btheta \mid \bx^*) \, d\btheta\,.
\]
This ratio can be estimated by Monte Carlo using the samples $(\btheta_i^{(\ell-1)})_{1 \leq i \leq N}$ from $\widehat{\pi}^{(l-1)}(\btheta \mid \bx^*)$ that were generated at iteration $\ell - 1$ of Algorithm \ref{algo:snle}, namely
\[
\widehat{R_\ell} = \frac{1}{N} \sum_{i=1}^N \frac{q^{(\ell)}(\bx^* \mid \btheta_i^{(\ell-1)})}{q^{(\ell-1)}(\bx^* \mid \btheta_i^{(\ell-1)})}\,.
\]
Letting 
$C_0 = 1$ and $q^{(0)}(\cdot \mid \btheta) \equiv 1$, we estimate the final evidence as the product:
\begin{equation}
\label{eq:SIS}
\widehat{C_L} = \prod_{\ell=1}^L \widehat{R}_\ell = \prod_{\ell=1}^L \left( \frac{1}{N} \sum_{i=1}^N \frac{q^{(\ell)}(\bx^* \mid \btheta_i^{(\ell-1)})}{q^{(\ell-1)}(\bx^* \mid \btheta_i^{(\ell-1)})} \right).
\end{equation}

This estimator can be computed directly from the sequence of density approximations and posterior samples produced by Algorithm \ref{algo:snle}. Conceptually, it draws inspiration from classical methods that estimate partition functions or marginal likelihoods by exploiting a sequence of intermediate distributions.

\begin{itemize}
\item The \textbf{Steppingstone Sampling (SS)} estimator~\citep{Xie2011} constructs a geometric path between the prior and posterior by introducing a series of tempered (power) posteriors. Each intermediate distribution bridges the gap between prior and posterior, allowing for stable importance weight updates across steps. The marginal likelihood is then obtained as a product of ratios of normalizing constants between consecutive steps, similar in spirit to the recursive product formulation in SIS-SNLE.

\item \textbf{Sequential Monte Carlo (SMC)} samplers~\citep{DelMoral2006} also construct a sequence of intermediate distributions and estimate normalizing constants using weighted particle approximations. These methods rely on importance sampling and resampling steps to control the variance and particle degeneracy across the sequence.
\end{itemize}

In our setting, the rounds of the SNLE algorithm naturally yield a sequence of increasingly accurate posterior approximations. As the neural likelihood surrogates are refined, the corresponding posterior approximations become more concentrated around high-probability regions. This structure mirrors the gradual transition seen in both SS and SMC approaches, making the Sequential Importance Sampling strategy a natural and computationally efficient tool for marginal likelihood estimation within the SNLE framework

\section{The IS-SNLE method}

The SIS-SNLE estimator \eqref{eq:SIS} described in the previous section leverages the intermediate approximations produced during SNLE, but it does not use MCMC samples from the final posterior approximation $\widehat{\pi}^{(L)}(\btheta \mid \bx^*)$ that is used for parameter estimation. This observation motivates a complementary strategy based on standard Importance Sampling (IS) \citep{Owen2013,Robert2004} using the final posterior approximation. Let $(\btheta_i^{(L)})_{1 \leq i \leq N}$ denote the MCMC sample from $\widehat{\pi}^{(L)}(\btheta \mid \bx^*)$. The IS-SNLE strategy proceeds in three steps:
\begin{itemize}
\item Train a neural density estimator $h(\btheta)$ on the MCMC sample $(\btheta_i^{(L)})_{1 \leq i \leq N}$ to approximate the marginal posterior $\widehat{\pi}^{(L)}(\btheta \mid \bx^*)$ 
\item Generate a new, independent and identically distributed (i.i.d.) sample $(\btheta_j^{(IS)})_{1 \leq i \leq N_{IS}}$
\item Estimate the marginal likelihood $C_L$ using the importance sampling estimator
\begin{equation}
\label{eq:IS}
\widehat{C_L} = \frac{1}{N_{IS}} \sum_{j=1}^{N_{IS}} q^{(L)}(\bx^* \mid \btheta_j^{IS}) \frac{\pi(\btheta_j^{IS})}{h(\btheta_j^{IS})}.
\end{equation}
\end{itemize}

Importance sampling can use any proper distribution as the sampling distribution, but would be optimal if one were able to use the posterior distribution. As it is unknown, we propose here to replace it by a neural estimate. This method hence fully exploits two crucial properties of neural density estimators using normalizing flows \citep{Papamakarios2021}: (i) the estimation $h(\btheta)$ is a proper density on the space of parameters $\Theta$, making it suitable for IS, and (ii) the very construction of $h(\btheta)$ allows efficient sampling from this distribution, so that we can generate the $N_{IS}$ independent draws easily and at a low computational cost.

This estimator requires only the final surrogate likelihood approximation $q^{(L)}(\bx^* \mid \btheta)$, the prior density $\pi(\btheta)$, and a tractable density estimate $h(\btheta)$ derived from the MCMC output. Crucially, the method is not specific to simulation-based inference and could be applied in any Bayesian setting where a posterior sample is available and a surrogate likelihood is accessible.

\section{The HM-SNLE method}

In addition to importance sampling-based approaches, one may consider the classical Harmonic Mean (HM) estimator for marginal likelihood \citep{Newton1994}. Following early work by \citet{McEwen2021}, \citet{Spurio2023} propose a retargeted version of this estimator tailored to simulation-based inference, which leverages the surrogate likelihood learned by SNLE. Remember that  $q^{(L)}(\bx^* \mid \btheta)$ denote the neural likelihood approximation after $L$ rounds and that $\pi^{(L)}(\btheta \mid \bx^*)$ denote the corresponding posterior. The marginal likelihood can then be expressed as:
\begin{equation*}
C_L^{-1} = \int \left\{\frac{\psi(\btheta)}{q^{(L)}(\bx^* \mid \btheta) \pi(\btheta)}\right\}
\pi^{(L)}(\btheta \mid \bx^*) \dd\btheta
\end{equation*}
where $\psi(\btheta)$ is any normalized density used to stabilize the estimator.
This design helps mitigate the notorious instability of the harmonic mean estimator, which otherwise suffers from infinite variance when the denominator becomes too small in the tails.

Just like the IS case, the optimal choice for $\psi(\btheta)$ would be the unknown posterior distribution.
\citet{Spurio2023} hence propose to estimate $\psi(\btheta)$ as a normalizing flow.
To avoid the double use of the data, the final MCMC sample from the posterior 
$\widehat{\pi}^{(L)}(\btheta \mid \bx^*)$ 
is split between an evaluating set $(\btheta_i^{(L)})_{1 \leq i \leq N_{\text{eval}}}$ 
and a learning set $(\btheta_i^{(L)})_{N_{\text{eval}+1} \leq i \leq N}$, 
with $1 \leq N_{\text{learn}} \leq N$ and $N_{\text{eval}} = N - N_{\text{learn}}$.
The learning set is used to to fit a NF to estimate $\psi(\btheta)$, and the 
evaluating set to 
fitted to a subset of the posterior samples generated via MCMC from  $\widehat{\pi}^{(L)}(\btheta \mid \bx^*)$. 
The final estimator is then 
\begin{equation}
\label{eq:HM}
\widehat{C_L^{-1}} = \frac{1}{N_{\text{eval}}} \sum_{i=1}^{N_{\text{eval}}} \left\{ \frac{\psi(\btheta_i^{(L)})}{q^{(L)}(\bx^* \mid \btheta_i^{(L)}) \pi(\btheta_i^{(L)})} \right\}.
\end{equation}

This estimator is simple to compute, leverages existing posterior samples, and avoids additional simulator calls. However, it remains sensitive to the tail behavior of the reweighting terms, and careful design of $\psi$ is essential—particularly in high-dimensional spaces where tail mismatch can still lead to unstable estimates.

\section{Concentrated or dilated normalizing flows}

It is well known in the literature that the HM estimator benefits from a proposal $\psi(\btheta)$ with lighter tails than the posterior.
To enforce such a behavior, \citet{Polanska2024} propose to use the very structure of the NF,
by manually changing the variance of the base distribution of the flow \textit{after it has been trained on the posterior sample}.
They apply a so-called multiplicative ``temperature'' $T$ on the variance of the base distribution
of the flow, with $0 < T \leq 1$.
The following transformation layers of the flow remain unchanged.
As the estimator~\eqref{eq:HM} is valid for any proper distribution, $\psi_{T}(\btheta)$
can then readily be used, in the formula, and provides for a more concentrated
instrumental distribution.

A similar trick can be used for our IS technique, with the crucial difference that the 
proposal distribution in this case must have heavy tails compared to the posterior distribution.
We hence propose to simply multiply the variance of the base distribution of the learned
flow by a temperature $T$, but with $T \geq 1$, so as to dilute the proposal distribution,
instead of concentrating it. The resulting NF $h_T(\btheta)$ is still a proper distribution
that is easy to sample from by construction.

\section{Numerical experiments}

We consider the simple Gaussian toy example described in \citet{Spurio2023}.
It is defined by the generative model
\[
\bx = (x_1, \ldots, x_d) \mid \btheta \sim \mathcal{N}(\btheta, I_d), \quad \text{with} \quad \btheta = (\theta_1, \ldots, \theta_d) \sim \mathcal{U}{[-2,2]}^d,
\]
and we fix the observed dataset to \( \bx^* = (0, \ldots, 0) \). In this setting, the marginal likelihood admits a closed-form expression
\[
C = \frac{1}{4^d (2\pi)^{d/2}} \int_{-2}^{2} \cdots \int_{-2}^{2} \exp\left(-\frac{1}{2} \sum_{i=1}^d \theta_i^2 \right) \, \dd \theta_1 \cdots \dd \theta_d = \frac{[\mathrm{erf}(\sqrt{2})]^d}{4^d},
\]
where $\mathrm{erf}$ denotes the Gauss error function \citep{Spurio2023}. The SNLE configuration is as follow
\begin{itemize}
\item $L = 5$ rounds, each with $N = 1{,}000$ simulations.
\item Likelihood model: Masked Autoregressive Flow (MAF) \citep{Papamakarios2017} with 5 transformations, each with 2 hidden layers of 64 neurons.
\item Early stopping after 20 epochs without validation improvement (10\% of data used for validation).
\item Posterior inference via slice sampling using 20 parallel chains, producing 1,000 approximate posterior samples.
\end{itemize}
The SNLE procedure is implemented using \texttt{zuko v1.4.0} \citep{rozet2022zuko} through the \texttt{sbi v0.24} interface \citet{Boelts2025}.
Note that, compared with \citet{Spurio2023} that used a SNLE with $10$ rounds of $10,000$ simulations each, our inference setup is much lighter, which illustrates the better efficiency of IS and SIS methods compared to HM.

For the Importance Sampling (IS) estimator, we generate $N=1,000$ samples from $\widehat\pi^{(L)}(\btheta \mid \bx^*)$ using the same slice sampler with 20 parallel chains, and train on this sample a MAF with 3 transformations, each with two hidden layers of 32 neurons.
We then vary the temperature from $1.0$ (no tempering) to $2$, taking $1.25$ as a default,
and generate $1,000$ new independent samples from the learned flow $h_T(\btheta)$.

For the Harmonic Mean (HM) estimator, we follow the setup of \citet{Spurio2023,Polanska2024}. 
We first generate $N'=2,000$ samples from $\widehat\pi^{(L)}(\btheta \mid \bx^*)$ using the same slice sampler with 20 parallel chains, and split this sample between a $1,000$ training set and
a $1,000$ evaluating set. We generated twice as many samples from the posterior compared to IS, so that the same number of $1,000$ samples can be used for training and evaluation in both cases.
The normalized density $\psi(\btheta)$ is then modeled using a Real-valued Non-Volume Preserving (RealNVP) \citep{Dinh2017} NF, and we apply a temperature $T$ varying between $0.5$ and $1.0$,
with a default to $0.8$.
This estimator is implemented using the \texttt{harmonic v1.2.3} package \citep{McEwen2021}.

As the inference method is stochastic, we re-ran all the algorithms $25$ times to account
for the variability of the results on a fixed dataset $\bx^*$.

\begin{figure}[H]
\begin{center}
\includegraphics[width=\linewidth]{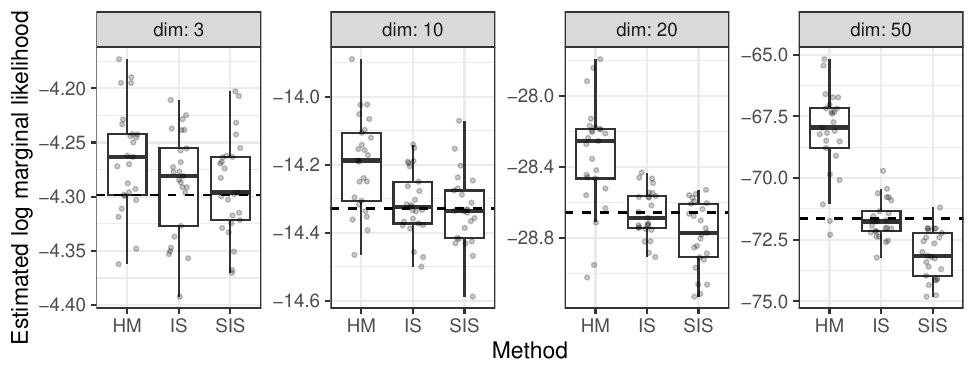}
\caption{Estimated log-marginal likelihoods with HM, IS and SIS. Each panel corresponds to a dimension $d$, and he dashed line corresponds to the true log-marginal likelihoods. For each method, $25$ independent estimates are shown, highlighting both accuracy and variability. We used a default temperature of, respectively, $0.8$ for HM and $1/0.8 = 1.25$ for IS.}
\label{fig:withHM}
\end{center}
\end{figure}

\begin{figure}[H]
\begin{center}
\includegraphics[width=\linewidth]{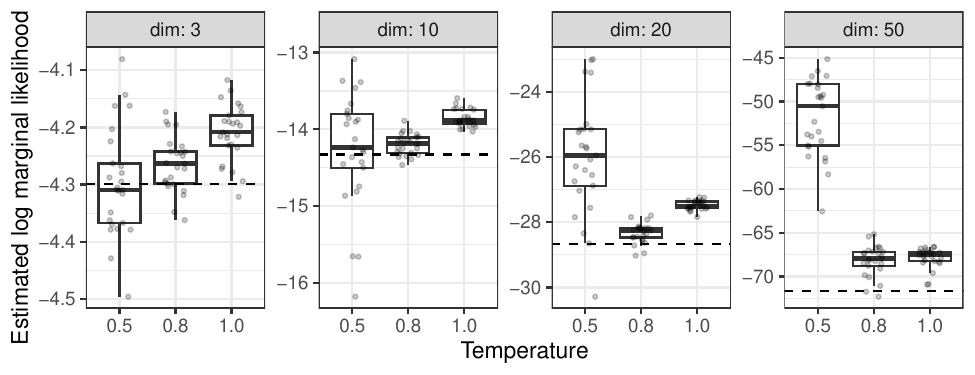}
\caption{Estimated log-marginal likelihoods with HM and varying temperature. Each panel corresponds to a dimension $d$, and he dashed line corresponds to the true log-marginal likelihoods. For each method, $25$ independent estimates are shown.}
\label{fig:onlyHM}
\end{center}
\end{figure}

\begin{figure}[H]
\begin{center}
\includegraphics[width=\linewidth]{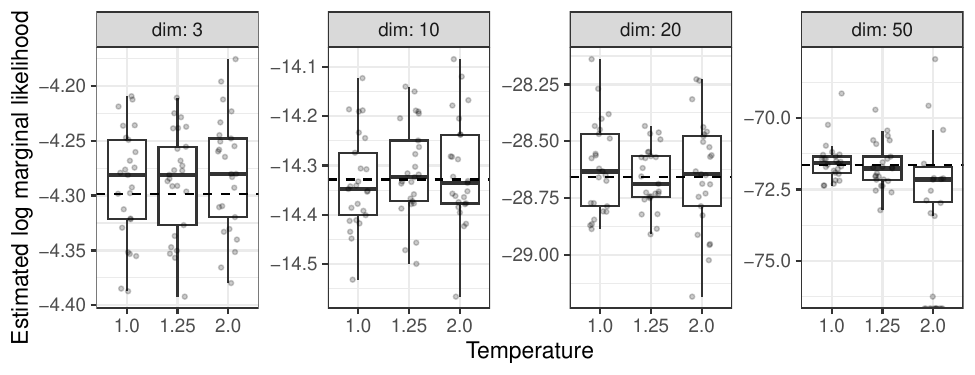}
\caption{Estimated log-marginal likelihoods with IS and varying temperature. Each panel corresponds to a dimension $d$, and he dashed line corresponds to the true log-marginal likelihoods. For each method, $25$ independent estimates are shown.}
\label{fig:onlyIS}
\end{center}
\end{figure}

Figure \ref{fig:withHM} illustrates that the HM estimator performed poorly as the dimensionality increased, exhibiting high variance and bias. The IS estimator consistently outperformed the SIS approach in higher-dimensional settings.
Figure~\ref{fig:onlyHM} and \ref{fig:onlyIS} show that the HM estimator was quite sensitive to the temperature parameter, while IS seemed to be more robust to variations of it.



\section{Discussion}

We introduced a general method to estimate the marginal likelihood in likelihood-free models using SNLE outputs. Our approach is simple, generic, and computationally efficient, leveraging the structure of SNLE to estimate the evidence via importance sampling. Challenges include potential variance due to support mismatch between the posterior and the proposal distribution, and the propagation of density estimation errors. Nonetheless, this provides a foundation for applying Bayesian model selection tools in the SBI setting. 
In addition to more extensive testing in more complex or realistic settings,
future research will investigate variance-reducing techniques, alternative proposal constructions, and robustness diagnostics. Extending this methodology to SNPE and SNRE is also a promising direction.

\bibliographystyle{plainnat}
\bibliography{snle_evidence}

\end{document}